\documentclass[aps,prb,showpacs,preprintnumbers,nofootinbib,twocolumn,10pt]{revtex4-2}
\usepackage{amsmath,amssymb}
\usepackage{bm}
\usepackage{tipa}
\usepackage{upgreek}
\usepackage{comment}
\usepackage{mathrsfs}
\usepackage{graphicx}
\usepackage{lipsum}
\usepackage{braket}
\usepackage{enumerate}
\usepackage{mathbbol}
\usepackage{booktabs}
\usepackage{gensymb}
\usepackage[normalem]{ulem}

\usepackage{color}

\usepackage[colorlinks=true,linkcolor=blue,allcolors=blue]{hyperref}

\def\equationautorefname~#1\null{Eq. (#1)\null}

\newcommand{\appref}[1]{\hyperref[#1]{App.~\ref*{#1}}}

\definecolor{purple}{rgb}{0.55,0.15,0.55}
\definecolor{nicegreen}{rgb}{0.28,0.75,0.19}

\usepackage{setspace}

\usepackage{bm}

\begin{document}

\title{Structure of Measurement-Induced Entanglement in Infinite-Randomness Critical States}
\author{Oliver Breach}
\email{oliver.breach@physics.ox.ac.uk}
\affiliation{Rudolf Peierls Centre for Theoretical Physics, Clarendon Laboratory, Parks Road, Oxford, OX1 3PU, UK}

\begin{abstract}

The impact of projective measurements on a many-body quantum state is tightly linked to its underlying entanglement structure. 
Critical states are particularly sensitive, as long-range entanglement allows local measurements to have global consequences.
This has been extensively studied in the context of critical states described by a conformal field theory (CFT), where measurements can alter critical properties in post-measurement quantities (`measurement-altered criticality') and induce long-range entanglement with universal features (`measurement-induced entanglement'). 
By contrast, little is known about the role of measurements on critical states with quenched disorder, which admit no CFT description.
Here, we develop a theory of measurements on one-dimensional critical states governed by infinite-randomness fixed points (IRFPs), focusing on two closely related examples: the random XXZ chain and the random transverse-field Ising model. 
We show that the measurement-induced entanglement decays as a power-law with universal exponent $(3-\sqrt{5})/2$ in both models, and that the minimum number of measurements required to establish this entanglement obeys a universal scaling form.
When only a finite density of measurements is made, we show that the critical properties of the state are remarkably robust: `measurement-altered criticality' is absent, with measurements either preserving the critical exponents, or destroying the criticality entirely.
These results establish IRFPs as an analytically tractable arena for measurements on critical states, complementary to the CFT case. The outcome randomness that necessitates replica methods in clean systems becomes trivial at infinite randomness, with the universal response governed by the statistics induced by the quenched disorder.

\end{abstract}

\maketitle

\section{Introduction}

The structure of entanglement in a many-body state has become a central concept in modern condensed matter theory. For example, the realization that local gapped Hamiltonians obey an area law for entanglement entropy \cite{hastings_area_2007,eisert_area_2010} has led to extensive development of both numerical and analytical techniques for interacting systems. Furthermore, entanglement structure underlies the modern understanding of phases of matter, in which two states are in the same phase if there exists a finite-depth quantum circuit consisting of local unitary gates that takes one state to the other \cite{chen_local_2010}. 

Modern quantum platforms possess the ability to apply not only unitary gates but also projective measurements. Unlike unitary gates, local quantum measurements can have a non-local effect on an entangled state, as is dramatically illustrated by quantum teleportation \cite{bennett_teleporting_1993,bouwmeester_experimental_1997}. 
This raises the question of what effects measurements can have on complex many-body states of matter. In particular, how is the impact of measurements on a state related to the initial entanglement?

This question has revealed a remarkable array of phenomena, including a close relation between phases of matter and the effect of measurements. 
One quantity which illustrates this is the measurement-induced entanglement (MIE), which captures the entanglement between two spatially separated regions of a state after the rest of the system has been measured in a local basis, averaged over all measurement outcomes \cite{lin_probing_2023,cheng_universal_2024}. 
MIE has significant operational meaning. For example, the difficulty of efficiently sampling from finite-depth quantum circuits in 2+1d can be interpreted as a consequence of the generation of finite long-range MIE \cite{mcginley_measurement_2025}, despite short-ranged entanglement before the measurements. Furthermore, MIE can be related to the sign problem: in sign problem free states the MIE is already upper-bounded by pre-existing mutual-information between the two regions \cite{lin_probing_2023,hastings_quantum_2015}.  Long-range MIE is also a necessary condition for a given state to be used for measurement-based quantum computation \cite{raussendorf_measurement-based_2003,briegel_measurement-based_2009,else_symmetry_2012,raussendorf_symmetry-protected_2017,raussendorf_computationally_2019}. 
Finally, optimizing MIE over local measurement protocols  yields the localizable entanglement, which upper bounds connected two-point correlation functions \cite{verstraete_entanglement_2004,popp_localizable_2005}.

Equally significant is the fact that MIE inherits a degree of universality from the underlying phase of matter. This is most striking in 1+1d, where the leading contributions to MIE are universal \cite{cheng_universal_2024}. For generic gapped phases, MIE decays exponentially with distance, while symmetry-protected topological phases have long-ranged MIE when measured in a symmetric basis \cite{else_symmetry_2012,stephen_computational_2017,tantivasadakarn_long-range_2024}. 
Gapless states are arguably the most interesting, since they have sufficient entanglement for long-range effects without relying on symmetries. Here, MIE is found to decay as a power-law, with a \textit{universal} exponent \cite{cheng_universal_2024}.  

Understanding the origin of this universality presents a substantial technical challenge due to the average over measurement outcomes weighted by the Born rule. 
The first analytic result for MIE in critical states was recently obtained for the free-boson conformal field theory (CFT) \cite{khanna_measurement_2026} by employing a replica trick and leveraging various free-boson techniques. Remarkably, the full probability distribution of post-measurement entanglement was shown to be universal \cite{khanna_universal_2025}.

In this paper we consider MIE in critical states which are not described by a CFT, whose structure is instead controlled by `infinite-randomness' fixed points (IRFPs). Examples of such systems include the 1d random transverse-field Ising model (rTFIM) \cite{fisher_critical_1995,fisher_random_1992} and the random antiferromagnetic XXZ model (rXXZ) \cite{fisher_random_1994}. Despite not having a CFT description, they still possess long-range entanglement with logarithmic scaling (once averaged over the quenched disorder). The rXXZ model was numerically shown in Ref.~\onlinecite{cheng_universal_2024} to have power-law decaying MIE; here, we develop an analytic understanding of this result, alongside MIE in the rTFIM.

As a complementary probe of the effect of measurements on states, we then turn to a different measurement scheme, in which measurements are made with finite density along the chain.  
Such measurements can lead to the phenomenon of measurement-altered criticality \cite{garratt_measurements_2023,weinstein_nonlocality_2023,murciano_measurement_2023,khanna_theory_2026,yang_entanglement_2023,lee_quantum_2023,paviglianiti_enhanced_2024,sala_quantum_2024,patil_highly_2024,liu_boundary_2025,naus_practical_2025,kumar_universal_2026}, where with suitable conditioning on measurement outcomes the critical properties of the states can be altered. 
We explore the impact of such measurements on the rTFIM and rXXZ and show that their critical properties are remarkably robust in contrast to their clean counterparts, which we attribute to the fact that the wavefunction is a product of `clusters', as is typical at IRFPs.

We note that rich behaviour has previously been observed from the \textit{dynamics} of measurements on systems involving disorder. For example, Ref.~\cite{zabalo_infinite-randomness_2023} studied a random unitary circuit interspersed with projective measurements with spatially varying probabilities, which led to a measurement-induced phase transition with strongly disordered Griffiths phases. 
Similarly, Ref.~\cite{tiwary_measurement_2026} considered Hamiltonian dynamics described by the XX spin chain, along with repeated measurements of $X_i$ and $Y_i$ with spatially varying rates, finding a new class of strongly disordered fixed points via SDRG methods. By contrast, here we focus on the properties of ground states of disordered systems, and make a single set of measurements on the system.

The outline of the paper is as follows. After defining MIE, we start by studying its behaviour in the rTFIM. Compared to the rXXZ which we discuss later, the effect of measurements is less drastic; although the MIE decays as a power-law, this entanglement was already present as mutual information before making the measurements.
Despite this, the underlying structure will provide important lessons for the structure of MIE in the rXXZ model, which we understand in two complementary ways: first, by relating the MIE to the entanglement across a half-system cut; and second, by identifying clusters of spins which act as independent systems for the measurements. 
In addition to analytically finding the exponent for the MIE, we will show that the minimum number of measurements required to induce this entanglement obeys a universal scaling form. 
Having understood MIE, we then consider measuring a finite density of sites, and study correlations and entanglement in the resulting states. Unlike the clean critical TFIM and XXZ model, their random partners are remarkably robust to measurement-altered criticality; either the critical properties remain fixed, or the measurements destroy criticality altogether.

\section{Measurement-Induced Entanglement}

\begin{figure}[h]
\centering
\includegraphics[width=7cm]{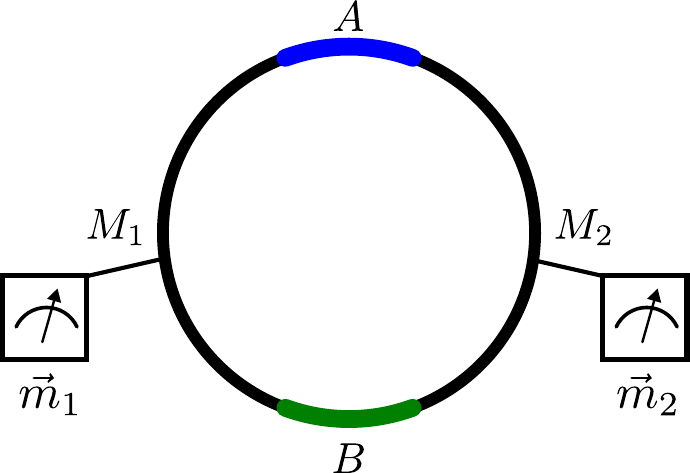}
\caption{Geometry of the setup for measurement-induced entanglement. Regions $A$ and $B$ are unmeasured; regions $M_1$ and $M_2$ are measured. We take the number of sites in regions $A$ and $B$ to be $|A|=|B|=m$ and the number of sites in the measured regions to be $|M_1|=|M_2|=r$. After the measurements, we study the entanglement entropy of regions $A$ and $B$, averaged over measurement outcomes $\vec{m}=(\vec{m}_1,\vec{m}_2)$.}
\label{fig:mie_geometry}
\end{figure}

Given a many-body quantum state $\ket{\psi}$, the measurement-induced entanglement (MIE) quantifies the entanglement between two separate regions A and B, when their complement $M=(AB)^c=M_1 \cup M_2$ is measured projectively. In this paper we use the geometry shown in Fig.~\ref{fig:mie_geometry}, in which we take periodic boundary conditions with $A$ and $B$ diametrically opposite, separated by measured regions $M_1$ and $M_2$.
We will consider the symmetric situation with $|A|=|B|=m$ and $|M_1|=|M_2|=r$.

Let $S_A(\ket{\psi})$ denote the entanglement entropy of region $A$, and denote outcomes of projective measurements on $M$ by $\vec{m}$. The MIE is defined as the entanglement entropy of the post-measurement state $S_A(\ket{\psi_{\vec{m}}})$, weighted over all measurement outcomes with the Born rule
\cite{cheng_universal_2024}:
\begin{equation}
\label{eq:mie_def}
    \mathrm{MIE}(A,B)[\ket{\psi}]=\sum_{\vec{m}} p_{\vec{m}} S_A(\ket{\psi_{\vec{m}}}).
\end{equation}
Since $M$ is measured projectively, and we here assume $\ket{\psi}$ to be pure, $S_A (\ket{\psi_{\vec{m}}})$ consists only of entanglement between regions $A$ and $B$.

\section{Random Transverse-Field Ising Model}
\label{sec:rtfim}

\subsection{Ground State Structure}

We first consider MIE in the random transverse-field Ising model (rTFIM), with Hamiltonian 
\begin{equation}
\label{eq:rtfim}
H = -\sum_i J_i Z_i Z_{i+1} - \sum_i h_i X_i,
\end{equation}
where $Z_i, X_i$ are the Pauli operators acting on site $i$.
The couplings $h_i, J_i$ are taken to be positive but random and drawn from the same distribution, placing the system at criticality \cite{fisher_critical_1995}. 

The low-energy properties of the rTFIM are well described by Fisher's strong-disorder renormalisation group (SDRG) \cite{fisher_random_1992,fisher_random_1994,fisher_critical_1995,dasgupta_low-temperature_1980}, which we now briefly review (more details can be found in App.~\ref{app:sdrg_review}). We start by assuming that the disorder is strong. Then if we choose the largest coupling in $\{J_i,h_i\}$, it will be much larger than all other couplings involving the same spin(s); we can therefore diagonalise this term in the Hamiltonian, and treat the rest of the Hamiltonian perturbatively. If the largest energy scale is $-J_i Z_i Z_{i+1}$, then diagonalising this term ensures that spins $i$ and $i+1$ point in the same direction, with the ground state lying in the subspace spanned by $\ket{\uparrow_i \uparrow_{i+1}}$, $\ket{\downarrow_i \downarrow_{i+1}}$. We call a set of spins thus constrained to point in the same direction a \textit{cluster}. Treating the remaining terms in the Hamiltonian to second order in perturbation theory, we find that this cluster experiences an effective transverse field of strength $h_{i,i+1} = h_i h_{i+1} / J_i$, with unaffected Ising couplings to its neighbours $J_{i-1}, J_{i+1}$. 
If, on the other hand, the largest energy scale is $-h_j X_j$, then we place spin $j$ in the $X$ direction $\frac{1}{\sqrt{2}}(\ket{\uparrow_j} + \ket{\downarrow_j})$; we call this `freezing' the spin. To second order in perturbation theory, fluctuations of this frozen spin lead to an effective Ising coupling between sites $j-1$ and $j+1$ of strength $J_{j-1,j+1} = J_{j-1} J_{j}/h_j$. 

\begin{figure}[h]
\centering
\includegraphics[width=7.6cm]{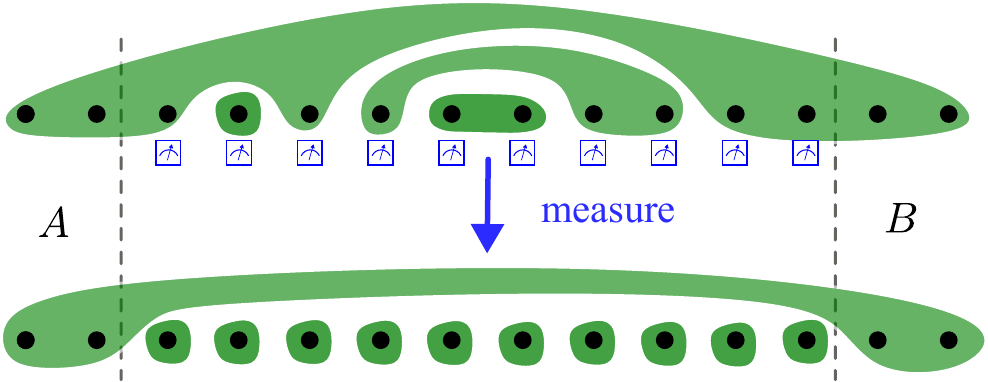}
\caption{An illustrative example of a ground state for the random transverse-field Ising model (top) and the result of $X$-basis measurements (bottom). Black dots illustrate sites with spin-$1/2$ particles. Clusters of aligned spins $\frac{1}{\sqrt{2}}(\ket{\uparrow\uparrow\dots \uparrow} + \ket{\downarrow\downarrow\dots \downarrow})$ are grouped together within the green regions. These clusters do not intersect. After measuring every site in the central region in the $X$-basis, there is measurement-induced entanglement $\ln{2}$ between regions $A$ and $B$. 
However, this entanglement was pre-existing in the sense of mutual information between regions $A$ and $B$. For ease of visualisation we use open boundary conditions here, instead of the geometry of Fig.~\ref{fig:mie_geometry}.}
\label{fig:rtfim_gs}
\end{figure}

Importantly, whether we couple two spins into a cluster or freeze a spin in the $X$ direction, the new Hamiltonian remains of the same form, except with one fewer spin and modified couplings (the effective magnetic moment of the spins may also be different, but this does not affect the Hamiltonian). As a result, this procedure can be iterated to find the ground state, which (asymptotically) consists of frozen clusters of spins on all length scales. A schematic illustration of the ground state for one disorder realisation is shown in Fig.~\ref{fig:rtfim_gs}. 
The validity of this approach relies on the fact that each iteration \textit{broadens} the disorder in the chain, such that the approach becomes asymptotically exact for describing the long-distance, low-energy physics.

To describe the evolution of couplings as the process is iterated, we can write flow equations for the probability distribution of bond couplings and field strengths. Let $\Omega = \mathrm{max} \{J_i, h_i \}$, and parametrise the couplings via $\zeta = \ln\frac{\Omega}{J_i}$ and fields via $\beta_i = \ln\frac{\Omega}{h_i}$. Taking the initial largest coupling to be $\Omega_0$, we introduce a renormalisation-group flow parameter $\Gamma = \ln\frac{\Omega_0}{\Omega}$. Then we denote the probability distribution of Ising couplings after iterating until a scale $\Gamma$ by $P_{\Gamma}(\zeta)$. We also seek the joint probability distribution $M_{\Gamma}(\beta, \mu)$ of fields with the effective magnetic moment $\mu$ (i.e. number of spins coupled together) of the cluster experiencing that field. The clusters described by this distribution, which have not yet been frozen, will be called `active'. 

As the highest energy scale is reduced and the RG flow parameter $\Gamma$ increases, we can write RG flow equations to describe the evolution of $P_{\Gamma}(\zeta)$ and $M_{\Gamma}(\beta,\mu)$ (see App.~\ref{app:sdrg_review}). Remarkably, one can show that the distributions flow to universal scaling forms
\begin{eqnarray}
\label{eq:prob_scaling}
P_{\Gamma}(\zeta) = \frac{1}{\Gamma} f\left(\frac{\zeta}{\Gamma}\right) \\
\label{eq:moment_scaling}
M_{\Gamma}(\beta,\mu) = \frac{1}{\Gamma^{1+\phi}} g\left(\frac{\beta}{\Gamma},\frac{\mu}{\Gamma^{\phi}}\right),
\end{eqnarray}
where $\phi = (1+\sqrt{5})/2$ is the golden ratio, with scaling functions $f(x)$ and $g(x)$. 

From these fixed-point solutions follow two important facts that we use below: (1) The density of active clusters at RG scale $\Gamma$ is $n_{\Gamma} \sim \Gamma^{-2}$, so the typical cluster size is $L \sim \Gamma^2$; (2) The typical number of spins (i.e. magnetic moment) in an active cluster at RG scale $\Gamma$ is $\mu \sim \Gamma^{\phi}$.

\subsection{Measurements in the X-basis}

Having established the structure of the ground state, we now consider the MIE. Recall the geometry in Fig.~\ref{fig:mie_geometry}. We will measure every spin in $M = M_1 \cup M_2$ in the $X$ basis, and then find the entanglement entropy between regions $A$ and $B$. 

In the previous subsection we saw that the ground state consists of clusters of (not necessarily contiguous) spins with wavefunction $\ket{\psi_c} = \frac{1}{\sqrt{2}} (\ket{\uparrow \uparrow \dots \uparrow} + \ket{\downarrow\downarrow \dots \downarrow}) $; the full ground state is a product state of such clusters. 
To see the effect of measurements, take a cluster, and divide the spins into three groups: those in $A$, those in $M$, and those in $B$. Make a projective measurement of $X_j$ for all $j \in M$. The post-measurement state for the unmeasured sites in $A\cup B$ is then\footnote{A quick way to see this is to consider the stabiliser for the initial state: $Z_1Z_2$, $Z_2Z_3$, $\dots$, $Z_{n-1}Z_n$, $X_1X_2\dots X_n$. Making a projective measurement of, say, $X_i$, $X_{i+1}$, $\dots$, $X_n$ means the new stabiliser is $Z_1Z_2$, $Z_2Z_3$, $\dots$, $Z_{i-2}Z_{i-1}$, $\pm X_1 X_2 \dots X_{i-1}$, $\pm X_{i}$, $\pm X_{i+1}$,  $\dots$, $\pm X_n$.} $(\ket{\uparrow \uparrow \dots \uparrow} \pm \ket{\downarrow\downarrow \dots \downarrow})/\sqrt{2}$, with the $\pm$ depending on the measurement outcomes (denoting the outcomes $x_j$, the sign is $\prod_j x_j$). If there is at least one spin in $A$ and at least one spin in $B$, the post-measurement entanglement entropy between $A$ and $B$ due to this cluster is $\ln{2}$, independent of measurement outcomes. 
Note however that before the measurement, the mutual information between $A$ and $B$ was already $\ln{2}$ --- this is necessitated by the fact that the ground state is sign-problem free \cite{lin_probing_2023}, and thus the MIE cannot exceed the pre-measurement mutual information.

The above argument shows that the measurement-induced entanglement is determined by the number of independent clusters which involve spins both in $A$ and in $B$; we will call such clusters `spanning'. 
While the number of spanning clusters is at most $m=|A|=|B|$, the number of spanning clusters which \textit{also} involve spins in region $M$ is at most two. 
This is because clusters are `non-crossing' (see Fig.~\ref{fig:rtfim_gs}): since clusters form by merging adjacent clusters, it is impossible to have two separate clusters involving spins in $A$, $M_1$, and $B$; the same applies to $M_2$. 
Furthermore, note that in the limit of large $r=|M_1|=|M_2|$, the probability that there is a single spanning cluster will decay more slowly than the probability of two spanning clusters, so we neglect these latter contributions to the MIE.   

The probability that regions $A$ and $B$ possess a spanning cluster can be heuristically determined as follows. 
In order for a spanning cluster to form, we require at least one spin in $A$ and one spin in $B$ to survive the decimation process (i.e. are not frozen) until the typical cluster size is $\sim r$; if this happens, then we expect that these spins will be in the same cluster. 
Since the typical number of spins per cluster is $\mu \sim \Gamma^{\phi}$, and the density of active clusters is $n_{\Gamma} \sim \Gamma^{-2}$, the density of active spins at scale $\Gamma$ is $\mu n_{\Gamma} \sim \Gamma^{(\phi-2)}$. Therefore, the probability that at least one spin is active in both $A$ and $B$ until the typical cluster size is $r\sim \Gamma^2$ is $p \sim \left[\Gamma^{(\phi-2)}\right]^2 \sim r^{-(2-\phi)}$. Thus the MIE, averaged over disorder realisations, decays as 
\begin{equation}
\overline{\mathrm{MIE}}  \sim r^{-\left(\frac{3-\sqrt{5}}{2}\right)}.
\end{equation}

The above scaling is the same as the scaling of the average spin-spin correlation function $C(r) = \overline{\langle Z_i Z_{i+r}\rangle}$. The average is dominated by rare events in which spins $i$ and $i+r$ are in the same cluster, in which case $\langle Z_i Z_{i+r}\rangle=1$. Since the MIE for large $r$ is dominated by cases where there is at most a single spanning cluster between $A$ and $B$, the scaling of average correlation functions and MIE is the same. Although we provided only a heuristic argument above, this behaviour of the correlation function can be more formally derived \cite{fisher_critical_1995}.

\section{Random-Singlet States}

\subsection{Ground State Structure}

We now turn to MIE for ground states $\ket{\psi(\{J_i\})}$ of the XXZ Hamiltonian
\begin{equation}
    H=\sum_i J_i\left( X_i X_{i+1} + Y_i Y_{i+1} + \Delta Z_i Z_{i+1} \right),
    \label{eq:hamiltonian}
\end{equation}
with positive random couplings $\{J_i\}$, and $-1/2 < \Delta \leq 1$. 
Like the rTFIM, the ground state can be found via an iterative RG transformation \cite{fisher_random_1994}. First the largest coupling $J_i\gg J_{i-1},J_{i+1}$ is identified, then we place the spins at $i$ and $i+1$ in a singlet. Quantum fluctuations of this singlet lead to an effective coupling between sites $i-1$ and $i+2$ of strength\footnote{This is correct for $\Delta=1$, but will differ by an overall factor for different $-1/2 < \Delta< 1$. This factor is, however, irrelevant in the RG flow and does not affect the low-energy, long-distance physics.} $J_{i-1,i+2} = J_{i-1}J_{i+1}/2J_i$. Iterating this procedure yields a ground state with non-crossing singlets on all length scales, as illustrated in Fig.~\ref{fig:rs_examples}(a). 
As this procedure is iterated, the probability distribution for the remaining bond strengths (in terms of $\zeta_i = \ln{\frac{\Omega}{J_i}}$ and $\Gamma = \ln{\frac{\Omega_0}{\Omega}}$) flows to the universal form 
\begin{equation}
P_{\Gamma}(\zeta) = \frac{1}{\Gamma} f\left( \frac{\zeta}{\Gamma} \right).
\end{equation}


\subsection{Measurements in Bell basis}
\label{sec:meas_in_bb}

\begin{figure}[h]
\centering
\includegraphics[width=7.6cm]{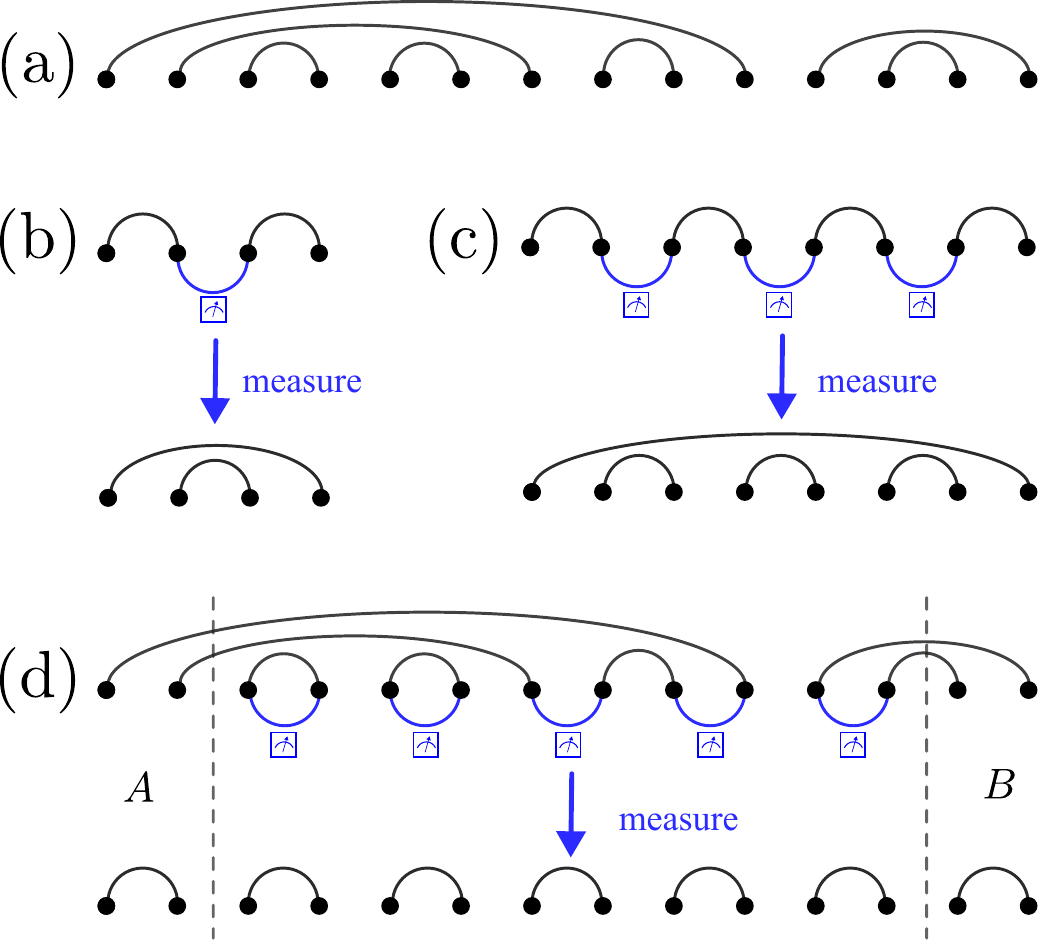}
\caption{In this figure we illustrate the effects of Bell-basis measurements on various initial states. In (a) we present an example random-singlet state. The black arcs curving upwards indicate Bell pairs between the corresponding sites. 
(b) demonstrates the basic premise of entanglement swapping, where blue curves below the sites indicate the location of Bell-basis measurements. In (c) we show how this extends naturally to a chain of Bell pairs, with the measurements inducing long-range entanglement between the ends of the chain.
In (d) we take an illustrative example of a random-singlet state and measure the central region. For this state, the measurement-induced entanglement is $0$. 
}
\label{fig:rs_examples}
\end{figure}

The singlet structure of the ground state means that single-site measurements in $M$ will not generate long-range entanglement. On the other hand, measurements in the Bell basis\footnote{We always take $|r|$ to be even, such that every site in $M$ is measured once.} lead to `entanglement swapping' \cite{zukowski_event_1993,pan_experimental_1998}, resulting in a non-local rearrangement of the original singlets (where we now have more general Bell pairs instead of only singlets). 
Fig.~\ref{fig:rs_examples}(b) illustrates the basic rearrangement on two Bell pairs, with  Fig.~\ref{fig:rs_examples}(c) showing how this can be chained together to produce long-range entanglement.
Bell-basis measurements in $M$ can thus lead to entanglement between $A$ and $B$ that was not present in the initial state. For these random-singlet states, the MIE counts the number of Bell pairs between $A$ and $B$ post-measurement. Note that pre-existing singlets between $A$ and $B$ are included in this count.

As discussed in Ref.~\onlinecite{cheng_universal_2024}, the MIE for random-singlet states from Bell-basis measurements is dramatically different for odd and even $m$. Post-measurement, regions $M_1$ and $M_2$ have been projected into nearest-neighbour Bell pairs. In the initial state every spin was in a singlet, and Bell-basis measurements can only rearrange the singlets, so if $m$ is odd there must be at least one post-measurement Bell pair between $A$ and $B$. Therefore the MIE, for large $r$, approaches $\ln{2}$ (even though the final answer is somewhat trivial, there is still non-trivial structure in how this entanglement is achieved; this will be discussed in Sec.~\ref{sec:num_meas}). For even $m$, by contrast, there can either be zero or finite MIE, with the relative probabilities dictating the decay with system size. 

In this section, we provide a detailed picture of MIE in random-singlet states, and provide an analytic calculation of its asymptotic power-law decay. We will first establish that the MIE is determined by chains of measured Bell pairs, then demonstrate that in the odd $m$ case there is always such a chain going from $A$ to $B$. For even $m$, we show that whether such chains exist depends on the half-system entanglement entropy, whose distribution can be computed directly. 

\begin{figure}[h]
\centering
\includegraphics[width=7.6cm]{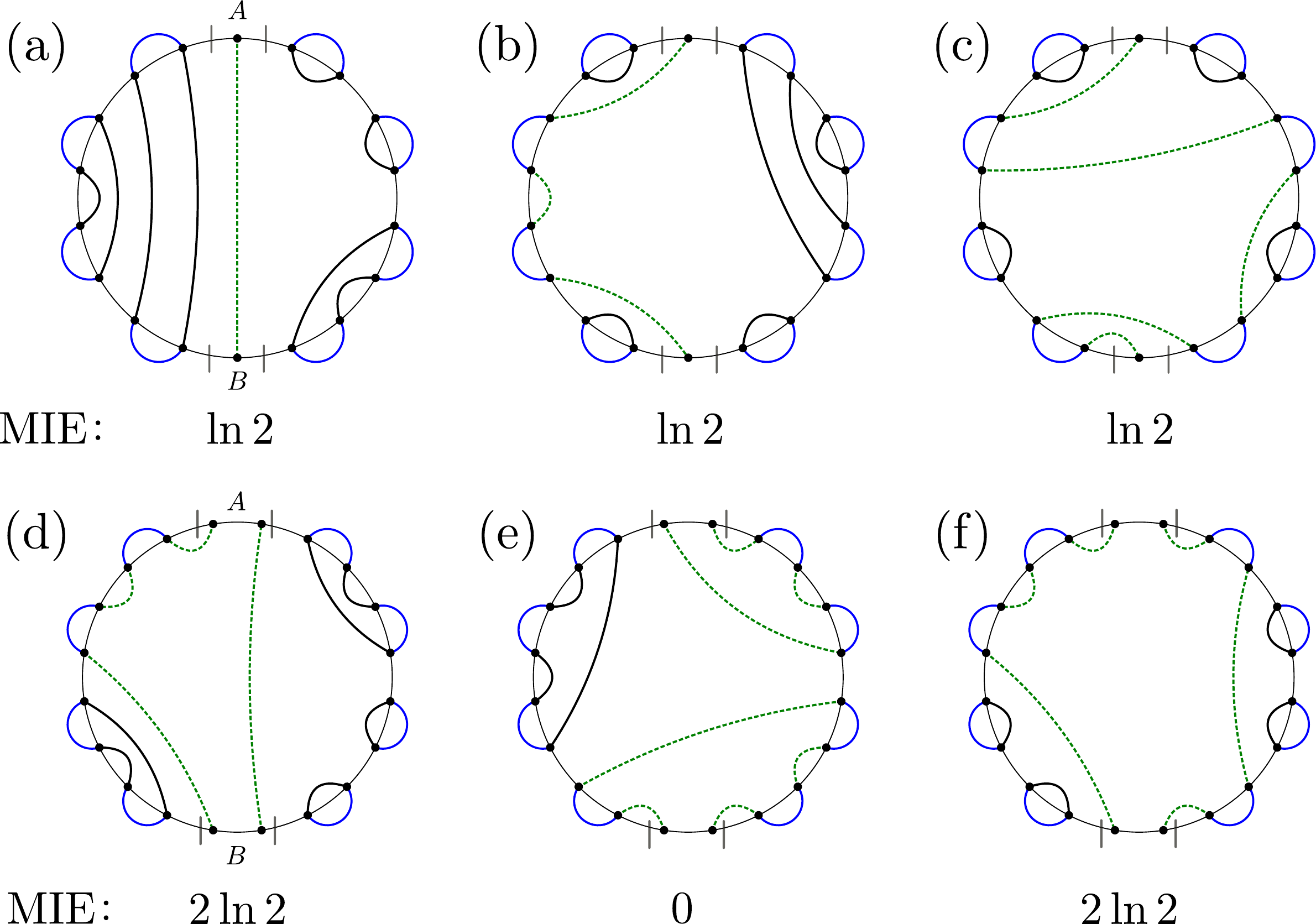}
\caption{We visualise the MIE by tracing paths drawn out by singlets with measurements represented by nearest-neighbour loops in $M$. The blue arcs on the outside represent measurement locations, and the interior arcs illustrate singlets in the initial state. The green (dashed) arcs are the singlets which participate in the creation of the post-measurement singlet; in the language of Sec.~\ref{sec:cluster_perspective}, this is the particular `measurement cluster' that involves spins in $A$ and therefore is relevant for MIE. Below each configuration, we state the MIE. 
For $m=1$ in (a)-(c), see that the path always leads from $A$ to $B$, due to the opposite `parity' of measurements in $M_1$ and $M_2$.  
For $m=2$ in (d)-(f), we obtain different post-measurement entanglement between $A$ and $B$ depending on the initial state. In particular, if a singlet crosses between $M_1$ and $M_2$ then there can be no MIE. The highlighted singlets here also correspond to the singlets in the `measurement cluster' which involves the sites in $B$. If this measurement cluster also involves sites in $A$, then the MIE is non-zero.}
\label{fig:mie_vis}
\end{figure}

Studying MIE for the random-singlet state with Bell-basis measurements has the attractive feature that it can be straightforwardly visualised. Given a spin in region $A$, the spin with which it is paired post-measurement can be found by (a) finding the spin with which it is initially paired in a singlet, then (b) moving to the neighbouring site with which this spin is measured, then (c) repeating steps (a) and (b) until arriving at an unmeasured site. If we connect the singlets and measured pairs with lines, this is equivalent to following the line of an unmeasured site until we end at another unmeasured site.
This is illustrated for simple chains of singlets in Fig.~\ref{fig:rs_examples}(b,c), and for an example random singlet configuration in Fig.~\ref{fig:rs_examples}(d).
If we start with a site in $A$ and end in $B$, then this pair of sites will contribute $\ln{2}$ to the MIE. 
This is illustrated in Fig.~\ref{fig:mie_vis} for several examples with $m=1$ and $m=2$.  
The measured pairs are represented by blue arcs around the exterior of the circle, with the initial ground-state singlets on the interior. The singlet pairs highlighted in green are those relevant for the MIE: following the pairings from the unmeasured region $A$ will either lead to the other unmeasured region $B$, in which case the measurement-induced entanglement is non-zero (a-d, f), or will lead back to $A$, indicating that the MIE is $0$ (e). 

From this visualisation, we can again understand the above statement that MIE is always at least $\ln{2}$ for odd $m$. 
For $m=1$, Figs.~\ref{fig:mie_vis}(a-c) illustrate that the line from the site in $A$ will always end at $B$. 
This is because when the line passes through sites in $M_1$ or $M_2$, the measurements always bring the path closer to $B$. This is ultimately due to the opposite parity of measurements in $M_1$ and $M_2$: after crossings between $M_1$ and $M_2$, illustrated in Fig.~\ref{fig:mie_vis}(c), the line continues to advance towards $B$.

Consider now $m=2$. The crucial difference to the $m=1$ case is that the `relative parity' of measurements in $M_1$ and $M_2$ is changed. To see this, consider Fig.~\ref{fig:mie_vis}(d-f). Starting from the left site of region $A$, follow the lines as before. If the line remains in $M_1$ (on the left), the parity of measurements is such that the line will always advance towards $B$, as illustrated in Fig.~\ref{fig:mie_vis}(d,f). In this case the MIE is $2\ln{2}$.  However, if the line crosses to the other side of the circle (from $M_1$ into $M_2$), then the parity of measurements will bring the line back to $A$, as shown in Fig.~\ref{fig:mie_vis}(e), and the MIE is $0$.

We now seek the relative probabilities of these outcomes over disorder realisations $\ket{\psi(\{J_i \})}$. 
The key is to notice that when the entanglement entropy between the two halves of the system (defined as a cut between the two sites in $A$ and the two sites in $B$) is non-zero, there can be no post-measurement entanglement, because the `path' must cross between the two halves of the system. Conversely, when the entanglement entropy is $0$, the MIE is non-zero.

The full probability distribution of the entanglement entropy across a cut was calculated in Ref.~\cite{devakul_probability_2017}. Applying this to a system of size $2(r+m)$ with periodic boundary conditions, and therefore two cuts, the probability that the entanglement entropy is $0$, scales as $P(S_{1/2}=0) \sim r^{-\alpha}$ with $\alpha = (3-\sqrt{5})/2 \approx 0.382$ (a derivation of this result is presented in App.~\ref{app:ee_distr}). 
Thus, the MIE averaged over disorder realizations scales as 
\begin{equation}
\overline{\mathrm{MIE}}  \sim  r^{-\left(\frac{3-\sqrt{5}}{2}\right)},
\end{equation}
just as in the rTFIM.

\begin{figure}[h]
\centering
\includegraphics[width=\linewidth]{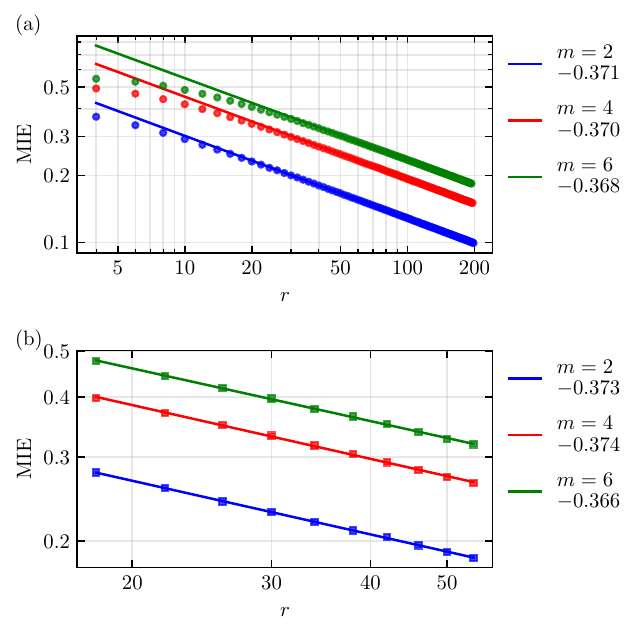}
\caption{Here we plot the disorder-averaged $\overline{\mathrm{MIE}}$ for the random-singlet state with unmeasured regions of size $m=2,4,6$. In (a) we use the strong-disorder renormalisation group approach, iteratively decimating bonds to find the ground state singlet structure, then keep track of how pairing is altered by the measurements.
In (b) we use free-fermion numerics for the random XX chain ($\Delta = 0$) as described in App.~\ref{app:ff}. The fitted exponents are shown on the right, consistent with the analytical prediction $(3-\sqrt{5})/2 \approx 0.382$.  
}
\label{fig:mie_data}
\end{figure}

This exponent is confirmed by numerically performing the decimations in the strong-disorder renormalisation group in Fig.~\ref{fig:mie_data}(a). To confirm that this scaling holds away from the exact fixed point, we also numerically compute the MIE using free-fermion numerics for the random XX chain ($\Delta = 0$) in Fig.~\ref{fig:mie_data}(b).

We now discuss the extension of the previous arguments for $m=1,2$ to general odd/even $m$. 
For $m>1$, there is the possibility to have more than one post-measurement singlet between $A$ and $B$. However, at most two Bell pairs can be \textit{induced} by the measurements (with one coming from measurements on each side), while all other Bell pairs must pre-exist. As a result, for odd $n>1$, the probability of MIE equalling $n\ln{2}$ scales as $\sim 1/r^{2(n-2)}$, and the leading term is the constant $\ln{2}$.

For even $m>2$, we can again have up to $m$ post-measurement Bell pairs between $A$ and $B$. The leading contribution remains two post-measurement Bell pairs, and the scaling of this contribution is unaltered from $m=2$ --- there is simply a combinatorial factor accounting for which sites in $A$ and $B$ can be paired. 
This is demonstrated numerically in Fig.~\ref{fig:mie_data}.
Again, two singlets is the most that can be \textit{induced} by the measurements, and the subleading corrections consisting of $n>2$ (even) post-measurement Bell pairs decay as $\sim \frac{1}{r^{2(n-2)}} \frac{1}{r^{2-\phi}}$.

\subsection{How many measurements are required?}
\label{sec:cluster_perspective}
\label{sec:num_meas}

An alternative way to arrive at the scaling of the MIE, which also allows for a finer understanding of \textit{how} the MIE is achieved, draws on our experience from the rTFIM. Expressing the rTFIM in the language of Majorana fermions produces a structure of paired Majoranas which is equivalent to the random-singlet state of the rXXZ chain. Physically, however, the `even' and `odd' bonds of this Majorana chain (with double the number of sites) have different meanings --- one set corresponds to the couplings $J_i$, while the other set corresponds to the fields $h_i$. Measurements of $X_i$, as we considered in Sec.~\ref{sec:rtfim}, also have the same effect as Bell-basis measurements on the random-singlet state, rearranging the paired Majoranas. 

This inspires an alternative perspective on the MIE in the rXXZ chain, in the case where $m$ and $r$ are even. Let $N=2m+2r$ be the total number of sites on the chain. The random singlet structure can be directly mapped onto a random paired-Majorana structure of the rTFIM, with $N$ Majorana fermions corresponding to $N/2 = m+r$ physical spins. The Bell-basis measurements are mapped to single-site $X$ measurements in the same geometry as in Sec.~\ref{sec:rtfim}, but with $|M_1|=|M_2| = r/2$ and $|A| = |B| = m/2$. Post-measurement entanglement in the size-$N/2$ rTFIM chain then corresponds to post-measurement entanglement in the size-$N$ rXXZ chain. Two paired Majoranas between $A$ and $B$ carry $\ln{2}$ entanglement entropy; this corresponds to $2\ln{2}$ entanglement entropy of two Bell pairs in the rXXZ case. Thus, the MIE in the rXXZ chain has the same scaling as the MIE in the rTFIM of $\sim r^{-(2-\phi)}$, despite the MIE for the rTFIM pre-existing as mutual information between $A$ and $B$. 

The above argument suggests introducing the notion of `measurement clusters' in the rXXZ chain, to provide more insight into the structure of the MIE. 
To this end, separate the bonds of the rXXZ chain into `even' and `odd'. With $m$ even, all measurements occur on bonds with the same parity --- for concreteness let us say on even bonds. 
Next, we define a `measurement cluster' as a set of sites which are connected through both measured bonds and singlets. These clusters are apparent in Fig.~\ref{fig:mie_vis} --- drawing both the singlets and measured bonds as arcs, such clusters form a closed loop. 
From the perspective of the clusters, singlets formed on even and odd bonds during the RG process have distinct roles. Singlets on odd bonds act to \textit{connect} clusters, joining together groups of sites. This is analogous to decimating a bond in the rTFIM. Singlets on even bonds, however, act to \textit{close} clusters, forming a closed loop, and preventing the cluster from growing large. This is analogous to field decimations in the rTFIM. 
The measurements can only rearrange singlets within these clusters. The MIE then counts the number of clusters which contain sites in both $A$ and $B$. 

\begin{figure}[h]
\centering
\includegraphics[width=7.6cm]{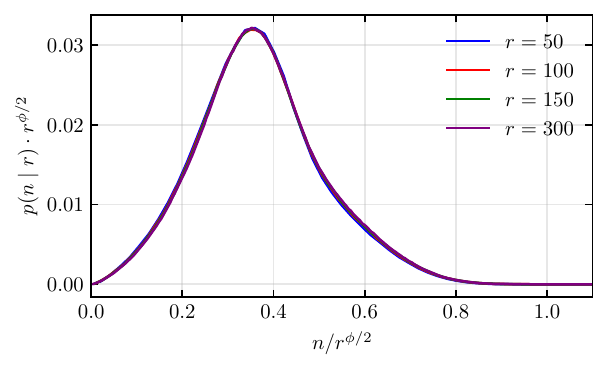}
\caption{
Here we show the scaling collapse for the distribution of the minimum number of measurements required to induce the non-zero MIE with system size $r$, in agreement with Eq.~\ref{eq:num_meas}. For even $m$, this is conditioned on there being non-zero MIE; for odd $m$, the MIE is always non-zero, and thus the probability is unconditional.}
\label{fig:num_meas}
\end{figure}

Thus, it remains only to find the scaling properties of these clusters. But this is already a solved problem --- the RG flow equations are identical to the flow equations for the rTFIM (Eqs.~\ref{eq:prob_scaling},~\ref{eq:moment_scaling}), where the magnetic moment is replaced with the number of sites in the cluster, the Ising couplings are replaced with the couplings on odd bonds, and the fields are replaced with the couplings on even bonds. 
From Eq.~\ref{eq:moment_scaling}, we further learn the scaling form for the distribution of the number $n$ of measurements required to induce the post-measurement entanglement --- i.e. what is the minimum number of Bell-basis measurements that one must make on the state to create the entanglement between $A$ and $B$. 
This number (conditioned on MIE being non-zero) obeys the universal scaling form
\begin{equation}
\label{eq:num_meas}
P(n|r) = \frac{1}{r^{\phi/2}} ~F\left( \frac{n}{r^{\phi/2}} \right)
\end{equation}
where we used the fact that the clusters at length scale $r$ form at RG flow parameter $\Gamma \sim \sqrt{r}$. We numerically demonstrate this scaling collapse in Fig.~\ref{fig:num_meas}.

A consistent definition of clusters across the entire chain requires $m$ to be even; otherwise, the parity of measured bonds in $M_1$ and $M_2$ are different. Despite this, the scaling form Eq.~\ref{eq:num_meas} still holds. This demonstrates that despite the `trivial' MIE in the odd $m$ case, there is still non-trivial structure in how this entanglement is generated.

We can understand the identical scaling for odd $m$ as follows. We can still define the notion of `measurement clusters' as groups of singlets which are closed systems under the measurements (equivalently, as connected loops in the visualisation). In the RG, these clusters grow and close with the same rates as in the even $m$ case. However, the `extra' site in regions $A$ and $B$ acts as a defect which means that one of the clusters containing spin(s) in these regions \textit{cannot close} until it spans the system. 
Despite this constraint that it cannot close, the independence of decimations means that it grows at the same rate as every other cluster. Thus, once this cluster has grown to reach the system size, the number of sites in this cluster will obey the universal scaling form Eq.~\ref{eq:num_meas}.

\section{Measurement-Altered Criticality}
\label{sec:mac}

We now turn to a different probe of the effects of measurements on many-body states. Instead of partitioning the system into regions, we treat every site equivalently, and on every site make a projective measurement with a finite probability $p$. We then study the non-local effect these measurements have on correlations and entanglement entropy.

Care must be taken when considering the quantities to study and how different measurement outcomes are handled \cite{garratt_measurements_2023,weinstein_nonlocality_2023}. For any quantity that is linear in the state's density matrix, such as a two-point correlation function between unmeasured sites $\langle Z_i Z_{j} \rangle$, averaging over measurement outcomes will lead to no change relative to its pre-measurement value. As a result, one should either consider averaging quantities which are non-linear in the density matrix, such as $\langle Z_i Z_{j} \rangle^2$ and entanglement-entropies, or study their value in particular `forced' measurement outcomes. 

For clean critical chains, such measurements lead to `measurement-altered criticality'. In the TFIM (and XX(Z) chain), measurements act as marginal operators in certain postselected ensembles, causing exponents of correlation functions to vary continuously with measurement strength/probability \cite{weinstein_nonlocality_2023,murciano_measurement_2023}. By contrast, in the Tomonaga-Luttinger liquid measurements can act as a relevant perturbation in the Born ensemble (for Luttinger parameter $K<1/2$), leading to multifractality in moments of correlation functions for the Born-averaged ensemble \cite{garratt_measurements_2023,khanna_theory_2026}.

In this section, we ask to what extent the \textit{random} TFIM or XXZ model display similar phenomena. We will show that the critical properties, such as averaged two-point correlation functions (and their moments), are remarkably robust to the measurements which preserve criticality. 
This absence of multifractality and varying exponents is a consequence of the binary nature of (average) correlations in the chain, where $\langle Z_i Z_j \rangle$ is either $1$ or $0$ depending on the disorder realisation.

We proceed by presenting some illustrative examples of choices of measurement bases, before discussing some general lessons about MAC at infinite-randomness fixed points.

\begin{figure*}
\centering
\vspace{-20pt}
\includegraphics[width=0.7\textwidth]{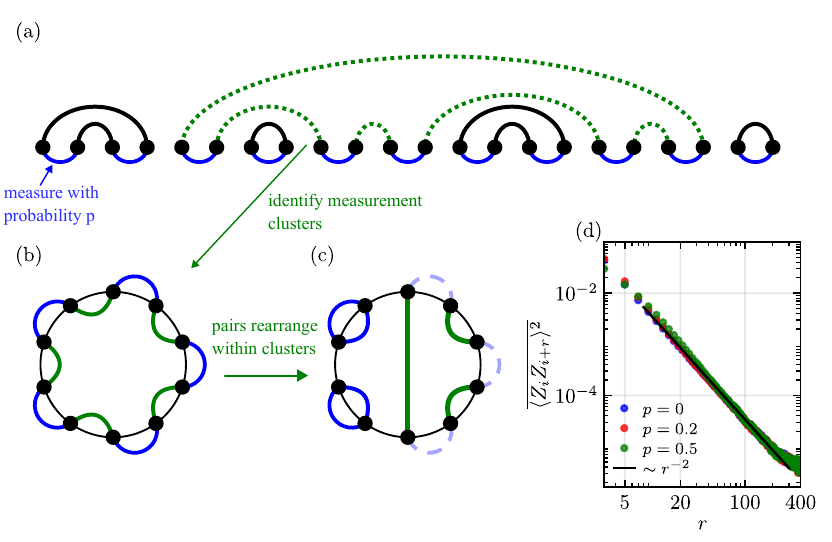}
\vspace{-15pt}
\caption{An illustration of the structure of Bell-basis measurements on a random-singlet state. In (a) we show a representative random-singlet state configuration. The blue arcs below illustrate the location of potential Bell-basis measurements --- each is made with a probability $p$. In green (dashed), we highlight singlets making up a single `measurement cluster', i.e. singlets which form a closed system under the measurements. This measurement cluster can be isolated as in (b), and forms an effective loop of nearest-neighbour singlets, which are then rearranged by the measurements as in (c). Since this rearrangement is local (see main text), the post-measurement correlations still decay as $r^{-2}$, as illustrated numerically using SDRG methods in (d).}
\label{fig:cluster_rearrangement}
\end{figure*}

\subsection{rTFIM}

It should by now be clear that, from the perspective of measurements, the key structural feature of the ground states of IRFPs is that they separate into `clusters' which act independently under measurements. This is clearest for the rTFIM with single-site measurements.

To start, consider making measurements of $X$ with probability $p$ on each site. As discussed in Sec.~\ref{sec:rtfim}, within a given cluster with initial wavefunction $\frac{1}{\sqrt{2}} (\ket{\uparrow \uparrow \dots \uparrow} + \ket{\downarrow \downarrow \dots \downarrow})$, $X$ measurements result in a wavefunction $\frac{1}{\sqrt{2}} (\ket{\uparrow \uparrow \dots \uparrow} \pm \ket{\downarrow \downarrow \dots \downarrow})$ on the unmeasured qubits. As a result, moments of correlation functions $\overline{\langle Z_i Z_j \rangle ^k}$ (where $\overline{\cdot}$ indicates both disorder and measurement-outcome averaging) are unmodified. Notice in particular the binary nature in which $|\langle Z_i Z_j \rangle| = 0,1$ is a direct consequence of the IRFP ground state structure. 
Such measurements do not affect the scaling of entanglement entropy (EE). Pre-measurement, the disorder averaged entanglement entropy scales as $S \sim \frac{\ln{2}}{6} \ln{L}$, due to clusters which involve sites both inside and outside the region of length $L$. To remove the contribution of a given cluster to the EE, all sites inside or all sites outside the region must be measured --- asymptotically this probability is exponentially suppressed as $p^{\mathcal{O}(L^{\phi/2})}$. 
All of these conclusions are independent of the particular measurement outcome, and therefore different behaviour is not seen in any particular forced outcome ensemble.

A different picture emerges under $Z$ measurements. A single $Z$ measurement collapses an entire cluster into $\ket{\uparrow \uparrow \dots \uparrow}$ or $\ket{\downarrow \downarrow \dots \downarrow}$, destroying connected correlations. Clusters of size $n \sim \Gamma^{\phi}$ survive with probability $(1-p)^n$, so for finite $p$ the large clusters which dominate the decay of connected correlation functions $\langle Z_i Z_j \rangle_c$ are collapsed beyond a finite length scale\footnote{The number of sites in a cluster is $n \sim L^{\phi/2}$. A cluster survives with probability $(1-p)^n$, so the lengthscale at which this becomes some finite fraction $c$ is $(1-p)^{\xi^{\phi/2}} \sim c$, which for small $p$ gives $\xi(p) \sim p^{-2/\phi}$.} $\xi(p) \sim p^{-2/\phi}$, and the entanglement entropy saturates at $S \sim \frac{\ln{2}}{6} \ln{\xi(p)}$. By contrast, note that the average $\overline{\langle Z_i Z_j \rangle^2} \to \mathcal{O}(1)$ at large $|i-j|$.


Another interesting possibility is measurements of $Z_i Z_{i+1}$, which act to tie together clusters. For instance, measuring $Z_2Z_3$ with outcome $+1$ on two nearest-neighbour clusters $\frac{1}{2} (\ket{\uparrow \uparrow} + \ket{\downarrow \downarrow})(\ket{\uparrow \uparrow} + \ket{\downarrow \downarrow})$ produces the state $\frac{1}{\sqrt{2}}(\ket{\uparrow \uparrow \uparrow \uparrow} + \ket{\downarrow \downarrow \downarrow \downarrow})$. Thus, $Z_i Z_{i+1}$ measurements can lead to larger clusters, enhancing $\overline{ \langle Z_i Z_j \rangle ^2 }$ correlations, since sites in clusters which have been joined by the measurements become correlated. 
However, for $p<1$ this joining does not modify the asymptotic scaling of the correlation functions. 
This is because only neighbouring/nested clusters can be correlated, and connecting a chain of $n$ clusters has exponentially decaying probability $p^n$.
Thus, clusters are only joined within a bounded range of scales, which is insufficient to modify the scaling of correlation functions.

\subsection{rXXZ}

We now turn to the rXXZ model, and first consider making Bell-basis measurements with probability $p$ on (for example) even bonds.\footnote{We do not choose to make Bell-basis measurements on every bond with this probability, since such measurements on adjacent bonds are non-commuting and therefore the order becomes relevant.} These Bell-basis measurements will act to rearrange the ground-state singlets via quantum teleportation. 
Thus, the effect of Bell-basis measurements might appear much more drastic than for the rTFIM above, but this is not the case.

It is useful to again adopt the idea of `measurement clusters' introduced in Sec.~\ref{sec:cluster_perspective}. For any given site, the only other sites it could be paired with after the measurement are sites in the same measurement cluster. 
Consider the effect of these measurements on the $\langle Z_i Z_j \rangle$ correlations. First note that the post-measurement pairing structure is independent of the particular outcomes; the outcome dependence is simply in which Bell pairs we have, and `measurement-induced' pairs will have random signs (but always magnitude $1$). Averaging over all outcomes leads to no change in $\langle Z_i Z_j \rangle$, but can produce changes in even moments such as $\langle Z_i Z_j \rangle^2$. 

To find $\overline{\langle Z_i Z_j \rangle^2}$, we need the probability that sites $i$ and $j$ are paired after the measurement. This requires first that $i$ and $j$ are in the same measurement cluster, and second
that the right set of bonds within the cluster are measured to create this pair (if a pair between $i$ and $j$ already exists, then no measurements are required).
Such a measurement cluster is illustrated in Fig.~\ref{fig:cluster_rearrangement}(a). As far as measurements are concerned, this cluster can be abstracted into a chain of nearest-neighbour singlets on periodic boundary conditions, with measurements made between the singlets with probability $p$ --- Fig.~\ref{fig:cluster_rearrangement}(b).
Crucially, measurements only have a \textit{local} effect on this nearest-neighbour chain. That is, the probability that a given site is in a pair of length $2j+1$ for $j\geq 1$ is $P(2j+1) = 2(1-p)^2 p^j$ (this comes from needing $j$ consecutive measurements to be made on one side of this site, with no measurements made either side). 
Such local rearrangements cannot modify the length distribution of singlets, and thus the critical properties remain unchanged.

We can similarly argue that entanglement entropy is unaffected by measurements. Consider some fixed region, and consider a singlet which crosses the boundary of this region, and hence contributes to the entanglement entropy. Although the probability that this particular singlet gets modified by the measurements is $p(2-p)$, the teleported singlet could still cross the boundary.
Again, this is clearest in the cluster perspective. For a singlet which crosses the boundary, consider its measurement cluster (which will necessarily involve another singlet crossing the boundary). The entanglement contribution from this cluster will be removed if and only if all sites on one side of the boundary are measured; otherwise, two Bell pairs from this cluster will still cross the boundary. By the same arguments as for clusters of the rTFIM, this possibility only gives an $\mathcal{O}(1)$ correction to the entanglement entropy.

As for the rTFIM, we can again turn to measurements which connect singlets instead of simply rearrange. For instance, $Z_i Z_{i+1}$ measurements on alternating bonds would connect singlets in a cluster into GHZ-like states, rather than rearranging the singlets. 
However, for the same reasons as for the rearrangement, this coupling is only local within the cluster.
The same argument as for the rTFIM again applies to making $Z_i Z_{i+1}$ measurements on every site with finite probability --- tying together clusters again cannot change the asymptotic scaling of correlation functions.

\subsection{General picture for MAC}

The emerging picture is that the critical properties of the IRFPs considered here are robust to broad classes of local measurements, as a consequence of the `clustering' property of their wavefunction. This clustering is a defining feature of IRFPs: the unbounded broadening of the coupling distribution (which leads to activated scaling with $z=\infty$) makes the real space RG which freezes the wavefunction into clusters asymptotically exact.
The state factorises into a product of clusters which are independent under measurements, and it is the statistical properties of these clusters which define the critical properties of the state.
Within these clusters, measurements at a finite density have only local effects (on a scale set by $p$), and therefore cannot alter the critical properties. 
The only way to meaningfully alter correlations is then to perform sufficiently destructive measurements such that, for example, connected correlations within a cluster are destroyed. 
Measurements which `connect' clusters similarly cannot affect the asymptotic properties, with the power-law scaling of the clusters winning out over the exponential decay associated with requiring multiple finite-probability measurements.
This robustness is also tied to a measurement outcome independence. 
Average correlations in the system are controlled by rare occurrences of strong correlation (e.g. $|\langle Z_i Z_j \rangle|$ in any disorder realisation is either approximately $0$ or $1$ in both systems considered), and this remains true post-measurement. Whether strong or weak correlations exist between a given pair of sites depends only on the measurement locations, and not the particular outcome --- the latter will only affect the sign of the correlations. As a result, no non-trivial physics arising from the Born probability rule can be observed: no particular postselected ensembles display unique behaviour, in contrast to the TFIM \cite{weinstein_nonlocality_2023}; and moments of correlation functions cannot display multifractality, in contrast to the Luttinger liquid \cite{khanna_theory_2026}.

\section{Discussion}

In this work, we have developed an analytic understanding of measurement-induced entanglement in random antiferromagnetic spin chains and the random transverse-field Ising model. For the rTFIM with measurements in the X-basis, we found that post-measurement entanglement required the existence of pre-measurement mutual information between the unmeasured regions $A$ and $B$. This arises when spins from both $A$ and $B$ are in the same `ferromagnetic cluster'; the probability of this occurring can then be computed using Fisher's strong-disorder renormalisation group, and is equivalent to the scaling of spin-spin correlations. For the random XXZ model, Bell-basis measurements have the effect of swapping the pairing of the original singlets via quantum teleportation. While the pre-measurement mutual information between $A$ and $B$ decays as $1/r^2$, we showed that the MIE decays as $1/r^{(2-\phi)}$ where $\phi=(1+\sqrt{5})/2$ is the golden ratio. This result can be understood either in terms of the entanglement structure of the initial state, or by adapting the notion of `clusters' from the rTFIM to describe groups of singlets which are linked by the measurement. 

These results complement recent analytic progress understanding the MIE in clean critical systems described by free-boson CFTs; it is interesting to contrast the two results.
In the random systems studied here, the MIE for a given disorder realisation is independent of the measurement outcomes; the outcome only determines the type of Bell pair (or cluster) carrying the entanglement. 
Then, when we average over disorder realisations, the average long-range MIE (for $m$ even) is dictated by the ratio of probabilities that the MIE is $2\ln{2}$ or the MIE is $0$ (in principle, the MIE can be as large as $m\ln{2}$, but these contributions decay faster with distance). 
In clean critical chains there is no disorder average, and the MIE depends on the measurement outcome. It is then interesting to consider the full probability distribution of measurement-induced entanglement over all outcomes, rather than looking only at the mean. Ref.~\cite{khanna_universal_2025} showed that this universal distribution is also bimodal, with peaks at $0$ and $\ln{2}$. Remarkably, the weight of the peak at $\ln{2}$ also controls the behaviour of the MIE, as in the disordered spin chains. 

We note that the structure of the state also lends itself to straightforward practical implementations on modern quantum platforms \cite{naus_practical_2025}. Knowing the disorder configuration, the singlet configuration can be efficiently computed using SDRG; using this information alongside the specific measurement outcomes would allow one to classically determine the post-measurement state efficiently. This knowledge could be used to verify the physical post-measurement entanglement.

In addition to understanding the MIE, we also investigated the possibility of measurement-altered criticality, and found that the critical properties of rTFIM and random XXZ model are remarkably robust to measuring with a finite density, in contrast to their clean counterparts \cite{weinstein_nonlocality_2023,murciano_measurement_2023}. This is a direct consequence of the clustering property that the wavefunctions of IRFPs display. Throughout, our analysis relied on the fixed-point wavefunction --- in the future it would be worthwhile considering the corrections to this picture, as well as considering the effect of measurements on \textit{typical} quantities, as opposed to rare-event dominated average quantities.

In this work, we considered arguably the simplest example of a disordered critical system --- the infinite-randomness fixed point of the one dimensional TFIM and XXZ model.
An important direction for future work is to consider broader classes of disordered critical points. 
A small step would be to consider the case of hyperuniform disorder, where one observes a line of infinite-randomness fixed points with continuously varying critical exponents \cite{crowley_quantum_2019}; since the ground state retains its cluster structure (albeit with different statistics), our methods should be directly applicable.
Alternatively, considering the case of random anyon chains \cite{bonesteel_infinite_2007,bonderson_measurement-only_2008,fidkowski_c-theorem_2008} could restore a measurement-outcome dependence that was absent here, and it would be interesting to see what role this plays in the physics. 
More drastically, one might consider finite-disorder critical points, such as the $XX$ chain with \textit{correlated} disorder \cite{alcaraz_random_2023} or in the superfluid-insulator transition for Luttinger liquids \cite{giamarchi_anderson_1988,altman_phase_2004,altman_superfluid_2010}. 
Finally, it would be interesting to consider the role that quenched disorder might play in the recently explored Bayesian critical points, which are the classical analogue of measurement-altered criticality \cite{nahum_bayesian_2025,putz_learning_2026}.

\begin{acknowledgments}

I am very grateful to Sid Parameswaran and Benedikt Placke for guidance and numerous discussions, particularly in the early stages of this project. I also acknowledge Romain Vasseur, John Chalker, and Akshat Pandey for useful discussions. 
I thank Sid Parameswaran, Calvin Hooper, James Walkling, and Ashley Wong for helpful comments on the manuscript.
I acknowledge support from a Leverhulme Trust International Professorship [Grant Number LIP-202-014], Merton College, and the Clarendon Fund. 
Lastly, I acknowledge the use of AI tools in the final stages of this project to produce code to plot the data and generate configurations to illustrate the arguments, and to improve the quality of text.
\end{acknowledgments}

\bibliography{rsbib}

\appendix

\section{Strong-Disorder Renormalisation Group}
\label{app:sdrg_review}

In this appendix we review the strong-disorder renormalisation group to find the ground state properties of the random XXZ model, then (primarily following Ref.~\cite{refael_entanglement_2004} and Ref.~\cite{devakul_probability_2017}) derive the probability that the half-system entanglement entropy is zero.

For simplicity we focus on the random Heisenberg chain 
\begin{equation}
H = \sum_i J_i \vec{S}_i \cdot \vec{S}_{i+1},
\end{equation}
but the results apply for more general $XXZ$ models $H=\sum_i J_i (X_i X_{i+1} + Y_i Y_{i+1} + \Delta Z_i Z_{i+1})$ for $-0.5 <\Delta \leq 1$. Here, the $\{J_i \}$ are drawn from some nonsingular distribution.

The core idea behind Fisher's strong-disorder renormalisation group \cite{fisher_random_1994} is to identify the strongest bond $J_i$, diagonalise this part of the Hamiltonian $H_0=J_i 
\vec{S}_i \cdot \vec{S}_{i+1}$, and treat the rest of the Hamiltonian perturbatively to second order. This leads to placing sites $i,i+1$ in a spin-singlet, with the new Hamiltonian on the remaining sites being a Heisenberg Hamiltonian with coupling between sites $i-1$ and $i+2$ of strength 
\begin{equation}
\label{eq:update}
\tilde{J}_{i-1,i+2} = \frac{J_{i-1} J_{i+1}}{2J_i}.
\end{equation}
This is justified if we assume that the disorder is strong, such that $J_i \gg J_{i+1}, J_{i-1}$. However, even if the initial disorder distribution is not broad, iterating this process broadens the distribution such that asymptotically the treatment is valid. 

To justify this statement, we can write down an equation describing the evolution of the probability distribution of couplings as we iterate the above process. We first reparametrise the coupling strength via $\zeta=\ln{\left(\frac{\Omega}{J} \right)}$, where $\Omega$ is the largest coupling $J_i$ in the chain. In terms of this parametrisation the update rule \ref{eq:update} becomes $\tilde{\zeta}_{i-1,i+2} = \zeta_{i-1} + \zeta_{i+1} + \ln{2}$; asymptotically this $\ln{2}$ can be neglected, so it will be dropped in the following. 
We also introduce an RG flow parameter $\Gamma = \ln{\left( \frac{\Omega_0}{\Omega}\right)}$, with $\Omega_0$ the initial value of $\Omega$. 

Let $P_{\Gamma}(\zeta)$ be the probability distribution for the couplings at RG flow parameter $\Gamma$. Consider increasing $\Gamma \to \Gamma + d\Gamma$. The probability distribution will change due to two effects. First, there is a redefinition of $\zeta \to \zeta - d\Gamma $ which shifts $P_{\Gamma}(\zeta)$, leading to $dP_{\Gamma}(\zeta) = \frac{\partial P_{\Gamma}}{\partial \zeta} d\Gamma$. This also accounts for the removal of the decimated bonds $0 \leq \zeta < d\Gamma$. Second, we add in new bonds with strength $\tilde{\zeta}_{i-1,i+2}$. The distribution for the couplings of such bonds is $\int d\zeta_1 d\zeta_2 \delta(\zeta_1 + \zeta_2 - \zeta) P_{\Gamma}(\zeta_1) P_{\Gamma}(\zeta_2)$, and the number of such bonds added is $P_{\Gamma}(0) d\Gamma$. We might worry that we should also be removing the neighbouring couplings to the decimated bond; however, these are simply drawn from the distribution $P_{\Gamma}(\zeta)$, and so would only lead to an overall scale factor accounting for the shortening of the chain. Renormalising the probability distribution therefore removes this effect. Putting these pieces together we find 
\begin{equation}
\label{eq:rg_flow}
\frac{dP_{\Gamma}(\zeta)}{d\Gamma} = \frac{\partial P_{\Gamma}}{\partial \zeta}+ P_{\Gamma}(0) \int d\zeta_1 d\zeta_2 \delta(\zeta_1 + \zeta_2 - \zeta) P_{\Gamma}(\zeta_1) P_{\Gamma}(\zeta_2).
\end{equation}

Fisher \cite{fisher_random_1994} showed that this equation has an (almost) universal attractor 
\begin{equation}
\label{eq:fp_distr}
P_{\Gamma}(\zeta) = \frac{1}{\Gamma} e^{-\zeta/\Gamma}
\end{equation}
which describes the long-distance, low-energy physics of the spin chain starting from almost any initial distribution of couplings.

Singlets which form at an RG time $\Gamma$ have a typical length given by 
$l\sim \Gamma^2$. This is because, using the fixed point distribution \ref{eq:fp_distr}, singlets are decimated at a rate $P_{\Gamma}(0)=1/\Gamma$, so the average density of surviving sites evolves as $dn/d\Gamma = -2n/\Gamma$ (the factor of $2$ is because two sites are removed in each decimation procedure). Thus, at large $\Gamma$, the density of remaining sites is $n(\Gamma) \propto 1/\Gamma^2$. The average length of singlets formed at the RG time $\Gamma$ is thus $l\sim \Gamma^2$.
The energy scale of these singlets formed at $\Gamma$ is $\Omega = \Omega_0 e^{-\Gamma}$. 
This gives us activated scaling of the energy gap (the lowest energy scale) $\log{\epsilon_{\mathrm{min}}^{-1}} \sim L^{1/2}$.

We can also verify the asymptotic correctness by noting that the typical coupling strength at $\Gamma$, from Eq.~\ref{eq:fp_distr}, is $\zeta_{\mathrm{typ}} \sim \Gamma$, i.e. $J_{\mathrm{typ}} \sim \frac{\Omega}{\Omega_0} \Omega$. Therefore as the largest energy scale is reduced to $\Omega$, the typical energy scale is reduced relative to this by a factor $\Omega/\Omega_0$.

\subsection{Entanglement Entropy Across a Cut}
\label{app:ee_distr}

We now focus on the entanglement entropy across a cut. The random singlet structure of the ground state means that, to compute the entanglement entropy, we need to compute the number of singlets which cross the cut. 

Focus on a particular bond $B$. To determine the rate of decimations across this bond (and hence rate of singlet formation across this bond), it is useful to introduce $Q_{\Gamma}(\zeta)$. This is the probability distribution for the coupling across this bond given that the last decimation event across this bond occurred at $\Gamma_0$, normalised such that $\int_0^{\infty}Q_{\Gamma}(\zeta) \equiv R(\Gamma,\Gamma_0)$ is the probability that this bond has not yet been decimated again since $\Gamma_0$. Knowing $Q_{\Gamma}(\zeta)$ gives us the rate of decimation, since this is just $Q_{\Gamma}(0)$. 

At $\Gamma=\Gamma_0$, immediately after a decimation across $B$, the bond strength distribution of $B$ is 
\begin{equation}
Q_{\Gamma_0}(\zeta) = \int d{\zeta_1} d\zeta_2 \delta(\zeta_1 + \zeta_2 - \zeta) P_{\Gamma_0}(\zeta_1) P_{\Gamma_0}(\zeta_2).
\end{equation}
Assuming the fixed point distribution for $P(\zeta)$, which will be asymptotically valid (i.e. there will be some entanglement formed across $B$ which won't be carried by singlets with this distribution, but this will be true at sufficiently long length scales, the entanglement for which will dominate over the short-range entanglement), we have $$Q(\zeta) = \frac{\zeta}{\Gamma_0^2} e^{-\zeta/\Gamma_0}.$$

As we run the RG subsequently, $Q_{\Gamma}(\zeta)$ satisfies a flow equation which has two contributions. First, we have a shift due to the redefinition of $\zeta$. Then we have the effect of decimating the bond to the left or to the right of $B$, each of which will lead to a new value of $\zeta$ at $B$. The removal of the `old' value of $\zeta$ leads to $dQ_{\Gamma}(\zeta) = -2P_{\Gamma}(0) Q_{\Gamma}(\zeta)$, whilst the addition of the `new' value of $\zeta$ leads to $dQ_{\Gamma} = 2P_{\Gamma}(0) \int d{\zeta_1} d\zeta_2 \delta(\zeta_1 + \zeta_2 - \zeta) P_{\Gamma}(\zeta_1) Q_{\Gamma}(\zeta_2)$. Here we have used the fact that the rate of decimations adjacent to $B$ is $2P_{\Gamma}(0)$, with the factor of $2$ coming from the $2$ sides.
This yields the flow equation
\begin{multline}
\frac{dQ_{\Gamma}(\zeta)}{d\Gamma} = \frac{\partial Q_{\Gamma}}{\partial \zeta}-2P_{\Gamma}(0) Q_{\Gamma}(\zeta)  + \\ 2P_{\Gamma}(0) \int d\zeta_1 d\zeta_2 \delta(\zeta_1 + \zeta_2 - \zeta) P_{\Gamma}(\zeta_1) Q_{\Gamma}(\zeta_2).
\end{multline}
In contrast to the flow equation for $P_{\Gamma}(\zeta)$, the normalisation is not fixed, with the probability of bond $B$ being decimated in a step $d\Gamma$ being $Q_{\Gamma}(0) d\Gamma = -dR(\Gamma,\Gamma_0)$. $R(\Gamma,\Gamma_0)$ is the probability of the bond not yet being decimated at $\Gamma$. 

Refael and Moore \cite{refael_entanglement_2004} solve this flow equation with the ansatz 
\begin{equation}
Q_{\Gamma}(\zeta) = \left( a_{\Gamma} + b_{\Gamma} \frac{\zeta}{\Gamma} \right) P_{\Gamma}(\zeta)
\end{equation}
with $a_{\Gamma_0}=0$ and $b_{\Gamma_0}=1$. This is natural since we expect that, if the bond survives a long time, it should `forget' the last decimation and approach the distribution $P_{\Gamma}(\zeta)$. 

Substituting the ansatz into the flow equation yields differential equations 
\begin{equation}
\Gamma \frac{da_{\Gamma}}{d\Gamma} = b_{\Gamma} - 2a_{\Gamma},~\Gamma \frac{db_{\Gamma}}{d\Gamma} = -b_{\Gamma} + a_{\Gamma}.
\end{equation}

The probability that no singlets form across the cut (possibly after some short RG time $\Gamma_0$ allowing for short-range entanglement) is given by $R(\Gamma,\Gamma_0) = \int_0^{\infty} d\zeta Q_{\Gamma}(\zeta) = a_{\Gamma} + b_{\Gamma}$. Solving the above differential equations we find 
\begin{multline}
\label{eq:survival_distr}
R(\Gamma,\Gamma_0) = \frac{1}{\sqrt{5}}\left[ \frac{3+\sqrt{5}}{2} \left(\frac{\Gamma_0}{\Gamma}\right) ^{(3-\sqrt{5})/2} - \right.  \\ \left. \frac{3-\sqrt{5}}{2} \left(\frac{\Gamma_0}{\Gamma}\right)^{(3+\sqrt{5})/2} \right].
\end{multline}

In the previous subsection, we saw that singlets formed at RG time $\Gamma$ have average length $l\sim \Gamma^2$. Therefore, for a finite chain of length $L$, we should stop the decimation procedure at $L\sim \Gamma^2$. (More generally one can show that the probability distribution for the final RG time for a finite system of size $L$ satisfies $p(\Gamma|L) = \frac{1}{\sqrt{L}} f(\Gamma/\sqrt{L})$ \cite{fisher_distributions_1998}. This is equivalent to $\Gamma = \sqrt{L} x$ with $x$ a random variable distributed as $f(x)$, and so we have $\ln{L} = 2\ln{\Gamma} - 2\ln{x} = 2\ln{\Gamma} + \mathcal{O}(1)$, and hence our substitution is valid in the large-$L$ limit.)

Therefore, for large $L$, the probability that no singlets (of length greater than some short-distance cutoff $L_0$) form across a cut in the middle of the chain scales as 
\begin{equation}
P(S=0) \sim \frac{1}{L^{(3-\sqrt{5})/4}}.
\end{equation}
In the main text, we consider periodic boundary conditions, and require that there is no entanglement between the two halves of the chain. We thus need there to be no entanglement across either of the two diametrically-opposite cuts, and therefore
\begin{equation}
P(S_{1/2}=0) \sim \frac{1}{L^{(3-\sqrt{5})/2}}.
\end{equation}

\subsection{Cluster Distributions}
\label{app:cluster_distribution}

In this subsection we discuss the notion of `active clusters' which can contribute to MIE at a given scale, and derive the scaling for the number of sites in an active cluster at a scale $\Gamma$. 
This is primarily a rephrasing of the results in Refs.~\cite{fisher_critical_1995,fisher_random_1994,fisher_random_1992} describing the magnetic moments of `spin clusters'.

Consider a system of size $L$, with Bell-basis measurements made on alternating bonds. We can think of these Bell-basis measurements as tying together singlets to produce long-range entanglement. We will call a set of singlets tied together by Bell-basis measurements a `cluster'.
Any set of singlets that are tied together to form a closed loop, however, will not contribute to the long-range entanglement; we call these `closed clusters'.
We seek the distribution of the number of singlets in `active clusters', namely clusters that have not formed a closed loop, at the scale $\Gamma \sim \sqrt{L}$. 

To find this distribution, we use the strong-disorder renormalisation group, but keep track of both the coupling strengths and the cluster sizes. To do this, we introduce separate notation for (logarithmic) couplings on even bonds, $\beta$, and odd bonds $\zeta$.  The even-bond couplings $\beta$ are on the bonds on which measurements take place. When a $\beta$ is decimated, this singlet will close a cluster. On the other hand, when a $\zeta$ is decimated to form a singlet, this will join together two open clusters.

Every surviving $\beta$ bond is associated to an active cluster. Let the number of `measurements' in this cluster be $\mu$. 
When a $\zeta$ bond is decimated, alongside the update rule for a new $\beta$, we join two active clusters together with size $\mu = \mu_1 + \mu_2$.

We therefore introduce the joint probability distribution $M_{\Gamma}(\beta, \mu)$, along with the probability distribution $P_{\Gamma}(\zeta)$. 
Now we consider increasing $\Gamma \to \Gamma+d\Gamma$, and look at how $M_{\Gamma}(\beta,\mu)$ changes. First, there is the shift due to the redefinition of $\beta \to \beta - d\Gamma$. Then there is the removal of $\beta$ bonds due to the decimation of $\zeta$ bonds. Each $\zeta$ decimation removes two $\beta$ bonds, and the number of $\zeta$ bonds decimated is $P_{\Gamma}(0) d\Gamma$; this gives $dM_{\Gamma}(\beta,\mu) = -2P_{\Gamma}(0) M_{\Gamma}(\beta,\mu)$. This decimation procedure creates a new $\beta$ bond with renormalised strength $\tilde{\beta} = \beta_1 + \beta_2$, and the corresponding clusters are merged together to give $\tilde{\mu} = \mu_1 + \mu_2$. We then need to normalise the new probability distribution via a change $d M_{\Gamma}(\beta,\mu) =  M_{\Gamma}(\beta,\mu) \left[ P_{\Gamma}(0) + M_{\Gamma}(\beta=0)\right]$, where $M_{\Gamma}(\beta) = \int d\mu M_{\Gamma}(\beta,\mu)$. Since the RG rules for coupling strengths are the same for odd and even sites, we have $M_{\Gamma}(\beta) = P_{\Gamma}(\beta)$. Therefore the normalisation term and the removal terms cancel, and flow equation for $M_{\Gamma}(\beta,\mu)$ is 
\begin{multline}
\label{eq:m_flow}
\frac{dM_{\Gamma}(\beta,\mu)}{d\Gamma} = \frac{\partial M_{\Gamma}(\beta,\mu)}{\partial \beta}  +  P_{\Gamma}(0) \int d\beta_1 d\beta_2 d\mu_1 d\mu_2 \big[ \\ \delta(\beta_1 + \beta_2 - \beta) \delta(\mu_1 + \mu_2 - \mu) M_{\Gamma}(\beta_1,\mu_1) M_{\Gamma}(\beta_2,\mu_2) \big]
\end{multline}
where $P_{\Gamma}(0)=\int d\mu M_{\Gamma}(0,\mu)$. Integrating over $\mu$ recovers Eq.~\ref{eq:rg_flow}.

Fisher \cite{fisher_random_1994} showed that this flow equation has a fixed point described by the scaling form 
\begin{equation}
M_{\Gamma}(\beta,\mu) = \frac{1}{\Gamma^{1+\phi}} Q\left(\frac{\beta}{\Gamma}, \frac{\mu}{\Gamma^{\phi}}\right)
\end{equation}
where $\phi = (1+\sqrt{5})/2$ is the golden ratio.
Integrating over $\beta$ finally yields the distribution of cluster sizes that we sought
\begin{equation}
p_{\Gamma}(\mu) = \frac{1}{\Gamma^{\phi}} F\left( \frac{\mu}{\Gamma^{\phi}}\right).
\end{equation}

\section{Free-Fermion Numerics}
\label{app:ff}

In this appendix we discuss the free-fermion numerics \cite{terhal_classical_2002,bravyi_lagrangian_2004} used to produce Fig.~\ref{fig:mie_data}(b). We closely follow the method described in Ref.~\cite{weinstein_nonlocality_2023}. 

The starting point is to make a Jordan-Wigner transformation, introducing Majorana fermions $\gamma_i$ satisfying $\{\gamma_j,\gamma_k\}=2\delta_{jk}$, where
\begin{eqnarray}
    i\gamma_{2j-1} \gamma_{2j}  = X_j 
    \\
    i\gamma_{2j} \gamma_{2j+1} = Z_j Z_{j+1}
    \\
    i\gamma_{2j-1} \gamma_{2j+2}=-Y_j Y_{j+1}.
\end{eqnarray}

The Hamiltonians for the rTFIM and rXX chain (i.e. $\Delta =0$, but swapping the roles of $Z_j Z_{j+1}$ and $X_j X_{j+1}$ in Eq.~\ref{eq:hamiltonian}) can then both be written in the form
\begin{equation}
    H=\frac{i}{4} \sum_{i,j=1}^{2N} \gamma_i A_{ij} \gamma_j,
\end{equation}
where $A_{ij}$ is a real antisymmetric matrix. This Hamiltonian can be straightforwardly block-diagonalised as 
\begin{equation}
    H=\frac{i}{2} \sum_{\alpha=1}^N \epsilon_{\alpha} \eta_{2\alpha-1}{\eta_{2\alpha}},
\end{equation}
where $\epsilon_{\alpha} \geq 0 $ and
\begin{equation}
    \eta_{\alpha}=\sum_{i=1}^{2N} R_{i\alpha} \gamma_i
\end{equation}
with $R\in SO(2N)$ such that the Majorana commutation relations are preserved. The eigenvalues can then be read off by setting $i \eta_{2\alpha - 1} \eta_{2\alpha} = \pm 1$, choosing all minus signs for the ground state. 

For these Gaussian states, all relevant information is contained in the two-point functions
\begin{eqnarray}
    G_{ij} = \langle i \gamma_i \gamma_j \rangle - i \delta_{ij}
    \\
    = \sum_{k,l=1}^{2N} R^T_{ik}R^T_{jl} \left[\langle i\eta_k \eta_l \rangle - i \delta_{lk} \right]
    \\
    = \sum_{\alpha} \left( R^T_{i,2\alpha} R^T_{j,2\alpha-1}-R^T_{i,2\alpha-1} R^T_{j,2\alpha} \right).
\end{eqnarray}
Higher-order correlation functions can be obtained by Wick's theorem, and the entanglement entropy of subregions can be found via Peschel's trick \cite{peschel_calculation_2003}.
Projective measurements of Gaussian operators $i\gamma_k \gamma_l=\pm 1$ (such as $Z_j Z_{j+1}=i\gamma_{2j} \gamma_{2j+1}$) can also be performed by applying the projection operator and using Wick's theorem to update the two-point correlation matrix
\begin{eqnarray}
    |\psi\rangle \mapsto P_{k \ell}^{ \pm}|\psi\rangle, \quad P_{k \ell}^{ \pm}=\frac{1 \pm i \gamma_k \gamma_{\ell}}{2}
    \\
    G_{i j} \mapsto G_{i j}^{\prime}=\frac{\left\langle\psi\left|P_{k \ell}^{ \pm} i \gamma_i \gamma_j P_{k \ell}^{ \pm}\right| \psi\right\rangle}{\left\langle\psi\left|P_{k \ell}^{ \pm}\right| \psi\right\rangle}.
\end{eqnarray}

We apply these techniques to obtain MIE in the random XX chain to produce Fig.~\ref{fig:mie_data}(b). For each value of $r$ and $m$, we take $4\times 10^4$ disorder realisations, where each $J_i$ is drawn from a power-law distribution with a constant shift: $J_i = 10^{\delta}\left(\frac{u_i}{\delta}\right)^{\delta} + 0.01$, where $u_i\sim U(0,1)$ and $\delta = 3$. Bell-basis measurements are implemented by measuring $Z_j Z_{j+1}$ and $Y_j Y_{j+1}$, both of which are Gaussian.
We found that under these projective measurements, numerical instabilities caused the Gaussian state to artificially lose its purity. We therefore, after each measurement, projected the state back onto a pure state by maintaining its eigenvectors but returning its eigenvalues to $\pm 1$. We confirmed on a subset of chains that this purification strategy, working with 64-bit numbers, produced equivalent results to using 256-bit numbers without the purification step. Using the latter precision it is not feasible to generate sufficient data to observe the MIE, hence the need for the purification.

\end{document}